\documentclass[prb,
superscriptaddress,showpacs,amsmath,amssymb,longbibliography]{revtex4-1}

\usepackage{xcolor}
\usepackage{graphicx}
\usepackage{hyperref}

\begin{document}

\title{Quantum and Classical mechanics vs QFT}

\author{G.E.~Volovik}
\affiliation{Landau Institute for Theoretical Physics, acad. Semyonov av., 1a, 142432,
Chernogolovka, Russia}

\newcommand{\nh}{\slash\!\!\!h}
\date{\today}

\begin{abstract}
15 years ago Dmitry Diakonov wrote the paper "Towards lattice-regularized Quantum Gravity", arXiv:1109.0091. \cite{Diakonov2011} In his approach, gravity with metric and tetrads arise from pre-geometric quantum fields leading to unusual dimensions of physical quantities. In particular, particle masses are dimensionless.\cite{Diakonov2012} We are trying to extend the Akama-Diakonov-Wetterich theory by introducing the Planck constants $\hbar$ and ${\slash\!\!\!h}=\hbar c$ as elements of the emergent metric. The inverse Planck constant $1/\hbar$ has the dimension of frequency, and, therefore, the mass $M$ of a particle, which has the dimension $\hbar\omega$, is dimensionless. In this extension, quantum mechanics emerges from the intrinsic quantum fields either in the symmetry breaking mechanism (GUT), or in the opposite mechanism of emergent symmetry in the low-energy corner (anti-GUT). In both cases, quantum mechanics (QM) serves as a bridge between the area of quantum fields (QFT) in the limit  $1/\hbar \rightarrow 0$, and the area of classical physics (CM) in the limit $\hbar \rightarrow 0$. In the GUT scheme the inverse Planck constants,  $1/\hbar$ and $1/{\slash\!\!\!h}$,  play the role of the order parameter of the symmetry breaking phase transition from the pre-geometric QFT state to the QM state, in which the quantum mechanics emerges together with the space-time metric.
In this phase transition, the integration over field variables in the QFT phase  transforms to a path integral formulation of QM, which in turn yields the laws of classical mechanics in the limit $1/\hbar \rightarrow \infty$. 
\end{abstract}
\pacs{
}
\maketitle 
 
 \tableofcontents
 
\section{Introduction}

In Akama-Diakonov-Wetterich (ADW) theory, the gravitational tetrads $e_\mu^a$ emerge as the bi-linear combinations of the Dirac quantum fields.\cite{Diakonov2011,Diakonov2012,Wetterich2021,Wetterich2022,Akama1978,Vergeles2025} 
This emergence of gravity can be considered in two different approaches, which can be referred to as the GUT and anti-GUT schemes.
 
The GUT scheme is based on symmetry breaking phase transitions. The original microscopic pre-geometric state has the highest possible symmetry, and with decreasing temperature the symmetry is spontaneously broken step by step. The first phase transition takes place at the super-Planck ultraviolet scale, where tetrads, metric, distances and geometry emerge from the original  state of the quantum fields. The further symmetry breaking transitions correspond to that, which leads to the Standard Model, with its symmetry breaking. 

Above the first phase transition, i.e. in the fully symmetric pre-geometric phase, there is no metric and thus no distance between the points. This may indicate that the correlations between points are either absent or, conversely, the same for any two points in the Universe. In both cases the original symmetry may include the symmetry under the relative space (and maybe time) translations. For example, the two point correlator $D({\bf r}_2,{\bf r}_1)$ is invariant not only under the conventional translation, ${\bf r}_1 \rightarrow {\bf r}_1 +{\bf a}$ and ${\bf r}_2 \rightarrow {\bf r}_2 +{\bf a}$, but is invariant under the independent translations, ${\bf r}_1 \rightarrow {\bf r}_1 +{\bf a}_1$ and  ${\bf r}_2 \rightarrow {\bf r}_2 +{\bf a}_2$. This is natural if the elements of the trans-Planckian substance are indistinguishable.

 In the  other approach -- in the lattice ADW theory, tetrads and distances continuously emerge in the low-energy corner of the ADW theory together with the corresponding symmetries. One obtains the quantum vacuum with emergent symmetries at low energy, where gravity emerges together with all the GUT symmetries. This scenario is similar to the so-called anti-GUT scenario, \cite{FroggattNielsen1991} which takes place also in the condensed matter systems. \cite{Volovik2003} At even lower energies the emerged GUT symmetries are spontaneously broken towards the Standard Model.

  In this anti-GUT lattice model of Diakonov theory, the original state is the full disorder, since the action describes the local four-fermion interaction, without any communications at large distances. Communications between the points of the lattice emerge in the low-energy (low-temperature) limit, where the coherence between the cells continuously appears and the propagating modes emerge. The effective distances emerge together with metric, which suggests two definitions of distance. One is the geometrical distance (the distance between the lattice points) and the other one is physical distance (the emergent dynamical distance).


In the theory with the vacuum as a plastic (malleable) fermionic crystalline medium -- the  “vacuum crystal”,\cite{KlinkhamerVolovik2019a} the geometric distance is obtained by counting the number of cells between the two lattice points. One has correspondingly the geometrical elasticity tetrads and the physical gravitational tetrads.

We make attempt to combine the  ADW theory and the approach in which the inverse Planck constants, $1/\hbar$ and $1/\nh \equiv 1/(\hbar c)$, correspond to the time and space components of the emergent Minkowski tetrads.\cite{Volovik2023,Volovik2023b}  In this approach, the Planck constants may appear in both scenarios: GUT and anti-GUT. In the GUT scenario,  $1/\hbar$ and $1/\nh$ emerge as the order parameters of the symmetry breaking phase transition, with $1/\hbar=1/\nh =0$ in the pre-geometric symmetric phase. This is opposite to the deterministic pre-geometric substrate when at high temperatures Planck constant goes to zero.\cite{Maiezza2026,Hossenfelder2013}.

\section{ADW theory and dimensions of physical quantities}
\subsection{ADW action}

In continuous GUT scenario of the ADW theory, the original action does not contain tetrads and metric and is described solely in terms of differential forms of quantum fields. The "cosmological term" in the action is
\begin{equation}
S_{\rm ADW}=\frac{1}{24}e^{\alpha\beta\mu\nu} e_{abcd} \int d^4x  \, \hat E^a_\alpha   \hat E^b_\beta \hat E^c_\mu \hat E^d_\nu \,,
\label{OriginalActions}
\end{equation}
and the Einstein-Cartan action is:\cite{Diakonov2011}
\begin{equation}
S_{\rm EC}=\frac{1}{4}e^{\alpha\beta\mu\nu} e_{abcd} \int d^4x  \, {\cal F}^{ab}_{\alpha\beta}
\hat E^c_\mu \hat E^d_\nu \,.
\label{EHC}
\end{equation}
Here the operator $ \hat E^a_\mu$ is:
\begin{equation}
 \hat E^a_\mu = \frac{1}{2}\left( \psi^\dagger \gamma^a\partial_\mu  \psi -  \psi^\dagger\overleftarrow{\partial_\mu}  \gamma^a\psi\right) \,,
\label{TetradsFermionsOperators}
\end{equation}
and ${\cal F}^{ab}_{\alpha\beta}$ is the Cartan curvature in terms of spin connection.

\subsection{Tetrads as order parameter of symmetry breaking}

In the ADW scenario, the gravitational tetrads $E^a_\mu$ are the vacuum expectation values of the bilinear fermion 1-form $\hat E^a_\mu$:
\begin{equation}
E^a_\mu=<\hat E^a_\mu>\,, 
\label{TetradsFermions}
\end{equation}
 while the metric field appears in the bilinear combination of the tetrad fields:
 \begin{equation}
G_{\mu\nu}=\eta_{ab}E^a_\mu E^b_\nu \,.
\label{DimMetric}
\end{equation}

In this scenario, tetrads emerge as the order parameter of the spontaneous symmetry breaking. While the original action is symmetric under both coordinate transformations and spin transformations, the order parameter is invariant only under combined transformations. The similar broken symmetry mechanism takes place in the B-phase of superfluid $^3$He, where the effective gravity vielbein also emerges as bilinear combinations of the fermionic fields.\cite{Volovik1990} 
Since gravity emerges from the matter fields, gravitational field can be considered as the type of the matter fields.\cite{Volovik2026t}  This allows us to give a different interpretation of Einstein equations, see Sec.\ref{GravFieldSection}.
 
The symmetry breaking scenario is the GUT-like version of the ADW theory.  In this version, symmetry breaking appears to be the general mechanism, which takes place in all the energy scales. First the tetrads and massless Dirac fermions emerge, and then the conventional GUT processes take place at lower energy leading to the Standard Model with its Higgs mechanism. The action for the emerging massless Dirac particles interacting with e.g. the scalar Higgs field is: 
\begin{equation}
S_{\rm Dirac}=\int d^4x\,  E\, (iE^\mu_a \bar\psi  \gamma^a \nabla_\mu \psi - \phi \bar\psi \psi)\,.
\label{Fermions}
\end{equation}

\subsection{Length dimensions of physical quantities}
 
As was emphasized by Diakonov,\cite{Diakonov2012} physical quantities in ADW theory have unconventional dimensions.
The  fermionic fields $\psi$ are dimensionless; tetrads $E^a_\mu$ are spacetime derivatives and thus have dimension 1/length  (we use $c=1$). Then the tetrad determinant $E$ has dimension $1/({\rm length})^4$; the metric $G_{\mu\nu}$ has dimension $1/({\rm length})^2$; the scalar boson field $\phi$ is dimensionless as well as the Dirac mass and cosmological constant.

\section{Extended ADW theory: Planck constants as elements of metric}

\subsection{Inverse Planck constants as the order parameter}

Typically,  $\hbar$ and $\nh$ are considered as fundamental constants. In such case one may take  $\hbar=\nh=1$ and forget about the Planck constants. Here we extend the ADW theory by considering the Planck constants  $\hbar$ and $\nh =\hbar c$ as  the elements of the metric of emergent Minkowski spacetime:\cite{Volovik2023}
\begin{equation}
G_{\mu\nu}^{\rm Mink}={\rm diag}(-1/\hbar^2,1/\nh^2, 1/\nh^2, 1/\nh^2) \,.
\label{Minkowski}
\end{equation}
In the symmetric state, the Planck constants are absent, $1/\hbar=1/\nh=0$. These inverse Planck constants, $1/\hbar$ and $1/\nh$, play the role of the order parameters of the symmetry breaking phase transition.

To make the further connection between the two approaches, let us assume that in the broken symmetry state, the tetrad determinant $E$ is constant, i.e. the emerging gravity is unimodular, and $E$ plays the role of the order parameter. For simplicity, let us take $c=1$ and thus $\nh =\hbar$. Then one has
\begin{equation}
E=-\frac{1}{\hbar^4} \,.
\label{determinant}
\end{equation}
In the symmetric pre-geometric phase, the tetrad determinant $E=0$, and thus $1/\hbar$ plays the role of the order parameter of the symmetry breaking phase transition from the QFT state to the state of quantum mechanics  (QM state) with its masses and trajectories, particles and fields, density matrix and collapse of wave function. The phase transition between the QM and QFT phases takes place when $1/\hbar=1/\nh=0$. 

In principle, there can be four different QM states, with different signs of $1/\hbar$ and $1/\nh$.

\subsection{Planck constant as coherence length}

Comparing the metric (\ref{Minkowski}) with that in Diakonov theory, one obtains that 
$\hbar$ and $\nh$ have dimension of time and length respectively.
In this approach, $\nh$ with its dimension of length, plays the role of the coherence length $\xi$ in the second order phase transitions.\cite{Volovik2023b}  The scale of $\xi$ is intermediate between the microscopic length scale $a$ and the macroscopic scale $l$, i.e. $a\ll\xi \equiv \nh \ll l$. In microscopic physics we use $a/\xi$ as small parameter (the QFT limit, $a \ll \nh$), while in macroscopic physics the small parameter is $\xi/l$ (semiclassical limit, $\nh \ll M l$).

The necessity of the small parameter for the emergent general relativity and/or gauge fields has been emphasized by Bjorken:\cite{Bjorken2003} "It would seem essential that there be a very small parameter which characterizes the violations", and "the emergence can only work if there is an extremely small expansion parameter in the game" (see also Ref. \cite{Volovik2008} where the lessons for quantum gravity from quantum hydrodynamics were discussed). In our case, the small parameter is $a/\nh \ll 1$, which is important for the emergence of quantum and classical mechanics, as well as classical gravity.

\section{Action in the QM phase and Planck length}

\subsection{Normalized units}

Using the normalization in terms of the dimensional Planck constant,
one can introduce the dimensionless tetrads and dimensionless metric, and correspondingly the dimensional fields (here we use $c=1$):
\begin{equation}
E_\mu^a  =\frac{1}{\hbar}  e_\mu^a \,,\,  \psi = \hbar^{3/2}\Psi  \,,\,  \phi = \hbar\Phi \,,\, E =\frac{e}{\hbar^4} \,,\, G_{\mu\nu}=\frac{g_{\mu\nu}}{\hbar^2}\,.
\label{normalization}
\end{equation}
Then the action in Eq. (\ref{Fermions}) transforms to the following action for Dirac fermions:
\begin{equation}
S=\hbar S_{\rm Dirac}=\int d^4x\,  e\, (ie^\mu_a \bar\Psi  \gamma^a  \hbar\nabla_\mu \Psi - M\bar\Psi \Psi)\,,
\label{Dirac}
\end{equation}
where $M=\hbar\Phi=\phi$ is the Dirac mass coming from the Higgs field $\Phi$. Eq.(\ref{Dirac}) remains valid even if the determinant $e$ is not kept constant.

The emergent mass $M$ has dimension of $\hbar\omega$, where $\omega$ is the corresponding frequency. That is why  $M$  is dimensionless both in the ADW approach and in its extension. 
This differs from the approach in which $\hbar$ is absent and mass has dimension of frequency.\cite{Ralston2012} In our case, $\hbar$ is the variable physical parameter.

Any action, including that in Eq. (\ref{Dirac}), has the dimension of length (or time), which is the same as the dimension of the interval. This demonstrates that the interval reflects dynamics rather than geometry.

\subsection{Planck length and length of Planck constant}

It is natural to assume that the value of $\hbar$ in the present low-temperature Universe, i.e. $\hbar(T=0)$, is on the order of Planck length, $l_{\rm P}=\sqrt{\hbar G}$, or vice versa, the Newton constant is determined by the present value of $\hbar$, i.e. $l_{\rm P}=\sqrt{\hbar G} \equiv \hbar$.  This also demonstrates that the Newton constant $G=l_{\rm P}^2/\hbar$ has dimension of length, $[G]=[L]$ (here again we use $c=1$).
In this case, the Planck mass, which is dimensionless as any mass, has the unit value:
\begin{equation}
M_{\rm P}=\sqrt{\frac{\hbar}{G}}=1\,.
\label{PlanckMass}
\end{equation}
This also follows from the Eq.(\ref{EHC}) for Einstein-Cartan term in the ADW theory.
Then all other masses $M$, which are also dimensionless, are actually expressed through the Planck mass, $m=M/M_{\rm P}=M$.

\subsection{Cosmological constant problem}

Let us also mention, that the main term in the ADW theory in Eq.(\ref{OriginalActions}) becomes the cosmological constant term with $\Lambda=M_{\rm P}^4=1$. This however does not contradict to the present value of the cosmological constant. Using the Hawking 4-form field to describe dynamics of the cosmological "constant", it was found that if the Universe has the natural value of 
$\Lambda=M_{\rm P}^4$ at the initial stage of the Big Bang, it will have the present value of the dark energy at present time.\cite{KlinkhamerVolovik2008}

\subsection{UV length scale}

On the other hand, the microscopic length scale $a$ corresponds to the intrinsic ultraviolet scale:
\begin{equation}
a \equiv l_{\rm UV}\ll l_{\rm P}=\hbar \,.
\label{scales}
\end{equation}

In principle, it is possible that this UV cut-off is applicable only for fermions, while for bosons the cut-off energy is much smaller. Such situation has been discussed on example of the anti-GUT analogue in Weyl superfluid.\cite{KlinkhamerVolovik2005}

The gravitational action does not contain the UV cut-off:
\begin{eqnarray}
S=\hbar S_{\rm EC}=\frac{1}{16\pi G}\int d^4 x \sqrt{-g}{\cal R}=
\label{GravityAction1}
\\
=\frac{\hbar}{16\pi} \int \frac{d^4 x}{l_{\rm P}^2}\sqrt{-g}{\cal R} \,.
\label{GravityAction2}
\end{eqnarray}

The same is for the action of the gauge fields.
For $\lambda_{\rm e} \gg r\gg l_{\rm P}$, where $\lambda_{\rm e}$ is the Compton wavelength of electron, one has the following zero-charge behaviour of the electromagnetic action:
\begin{equation}
S \sim  \hbar  \int d^4 x \sqrt{-g}  \ln \left(\frac{r^2}{l_{\rm P}^2} \right) F_{\mu\nu}F^{\mu\nu} \,.
\label{EMAction}
\end{equation}
Here the fields $A_\mu$ and $F_{\mu\nu}$ come from the gauging of the $U(1)$ field. Thus, being the geometric quantities, $A_\mu$ and $F_{\mu\nu}$ have  dimensions of $1/{\rm length}$ and $1/({\rm length})^2$ correspondingly. 

In principle, this action may contain the fermionic UV cut-off.\cite{KlinkhamerVolovik2005} 
Then one would have the extra contribution:
\begin{equation}
S \sim  \hbar  \int d^4 x \sqrt{-g}  \ln \left(\frac{l_{\rm P}^2}{l_{\rm UV}^2} \right) F_{\mu\nu}F^{\mu\nu} \,.
\label{EMActionUV}
\end{equation}
This allows us to connect the fermionic UV cut-off $E_{\rm UV}$ and thus the microscopic length scale $a$ with the number of generations of fermions,\cite{KlinkhamerVolovik2005} 
\begin{equation}
 \ln \left(\frac{E_{\rm UV}^2}{M_Z^2} \right) \sim \frac{580}{N_F} \,,
\label{alpha}
\end{equation}
where $M_Z$ is the $Z$-boson mass, which characterizes the electroweak energy scale. According to the Diakonov theory,\cite{Diakonov2011} the natural value of the number of the fermionic families is $N_F=4$, see Sec.\ref{open}.

  \section{Towards pure QFT in symmetric state}
  
From Eq.(\ref{EMAction}) it follows that approaching the symmetry breaking phase transition from the QM phase to the QFT phase, $1/\hbar \rightarrow 0$, the electric charge of electron is nullified. In conventional approach this corresponds to $\alpha =e^2/\hbar c \rightarrow 0$ as $\hbar \rightarrow \infty$. Gravity also disappears in this limit, as follows from Eq.(\ref{GravityAction2}). The limit $\hbar \rightarrow \infty$ corresponds to $G\rightarrow 0$. 

 Correlator of massless scalar field $\Phi=\phi/\hbar$ is
\begin{eqnarray}
D({\bf r})=\left<{\rm vac}|\Phi(0)\Phi({\bf r})|{\rm vac}\right>= \frac{1}{4\pi^2 r^2} \,.
\label{Corr}
\end{eqnarray}
It does not contain $\hbar$.
On the other hand, the correlator of the original dimensionless field $\phi$ is
\begin{eqnarray}
D({\bf r})=\left<{\rm vac}|\phi(0)\phi({\bf r})|{\rm vac}\right>= \frac{\hbar^2}{4\pi^2 r^2} \,.
\label{CorrOrig}
\end{eqnarray}
For the dimensionless field $\psi$ of massless fermions one has $D({\bf r}) \sim \hbar^3/r^3$, and for massive fields the exponential term is added, $\exp({-Mr/\hbar})$.

The correlator (\ref{CorrOrig}) depends on which limit is taken first, 
 $\hbar  \rightarrow \infty$ or  $r \rightarrow \infty$. 
 The Eq. (\ref{CorrOrig}) is valid only for distances larger than the coherence length, $r \gg \hbar$. 
 However, for distances below the coherence length, at $r<\hbar$, the correlator does not depend on $r$, $D(r<\hbar)  \sim 1$.  This remains valid for $\hbar \rightarrow \infty$ which may provide a clue as to what is happening in the symmetric phase.
 
 If in symmetric phase, $D$ at infinity does approach the constant value,  $D(r \rightarrow \infty) \rightarrow {\rm Const} \sim 1$, this would mean the action at a distance in the QFT phase. Then the action at a distance and non-locality will be automatically extended to the QM phase, i.e. to the broken symmetry phase where quantum mechanics emerges.

   \section{Wilczek theory}

Let us consider another pre-geometry model,\cite{Wilczek1998,Addazi2025,Addazi2025b,Meluccio2025} which is close to the ADW theory. Now instead of the bi-linear combination of the fermionic fields the vielbein emerges from the scalar fields $\phi^A$. This theory necessarily uses $SO(1,4)$ group in spin space (index $A=(0,1,2,3,4)$), while  the group in coordinate space is $SO(1,3)$  (index $\mu=(0,1,2,3)$). As in the ADW theory, the action is also expressed solely in terms of differential forms:
\begin{equation}
S_{\rm W}=\int d^4 x e^{\mu\nu\alpha\beta} e_{ABCDE}  \nabla_\mu \phi^A
\nabla_\nu \phi^B \nabla_\alpha \phi^C \nabla_\beta \phi^D \phi^E \,.
\label{funfbeinBosons}
\end{equation}
Here the scalar fields and the action are dimensionless as in the ADW aproach.
In condensed matter such dimensionless variables describe elasticity theory of crystals and quasicrystals.\cite{NissinenVolovik2018} Here 
$\phi^A(x^\mu)= 2\pi n_A$ with integer $n_A$ describe the crystal planes or the  quasicrystalline surfaces.

The term in the action, which includes massless Dirac fermions, is:
\begin{equation}
\int d^4 x e^{\mu\nu\alpha\beta} e_{ABCDE} ({\bar \Psi}\Gamma^A \nabla_\mu \Psi)
\nabla_\nu \phi^B \nabla_\alpha \phi^C \nabla_\beta \phi^D \phi^E \,.
\label{funfbein}
\end{equation}
Here $\Gamma^A$ are five Dirac $\gamma$-matrices $4\times 4$, and $\nabla_\nu$ includes the corresponding gauge field  --  the spin connection $A^{AB}_\mu$. 
Fermionic and bosonic fields are dimensionless, and the actions do not contain gravity with its tetrads and the Planck constants.  

Let us consider for simplicity only the bosonic action (\ref{funfbeinBosons}). Variation of the action has solution:
\begin{eqnarray}
\phi^A(x^\mu)=C \delta^A_4
\,.
\label{phiConst}
\end{eqnarray}
As a result, the tetrad order parameter appears in the broken symmetry phase as vacuum expectation value, 
\begin{equation}
E^A_\mu=\nabla_\mu \phi^A=A^{A4}_\mu\,.
\label{funfbein}
\end{equation}
Note that the 4-th component of vielbein is absent, $E^4_\mu=0$, and thus the $5\times 4$ vielbein transforms to tetrad. In a different way, the rectangular $5\times 4$ vielbein (f\"unfbein) was used in the Dirac spinor theory.\cite{ObukhovVolovik2024}

Since the tetrads $E^A_\mu$ are determined by the spin connection, they again have the dimension of inverse length, in the same way as in ADW theory.
Correspondingly, we can introduce $1/\hbar$ and $1/\nh$ as elements of $E^A_\mu$ in Minkowski spacetime, and one has
\begin{equation}
 E^0_0=\frac{1}{\hbar} \,.
\label{funfbeinMink}
\end{equation}

Eq.(\ref{phiConst}) gives the asymptotic values for the correlators in the symmetric phase at $r\rightarrow \infty$:
\begin{equation}
D(r\rightarrow \infty)=\eta_{AB}<{\rm vac}|\phi^A(0)\phi^B(r\rightarrow \infty)|{\rm vac}> =C^2\,.
\label{ConstCorrelatorAsympt2}
\end{equation}
This actually suggests the same correlations between all the points in space
\begin{eqnarray}
D({\bf r})=D(0)=D(\infty)\,.
\label{ConstCorrelator}
\end{eqnarray}
This is reasonable, since the length scale $\nh$ does not exist in symmetric phase, and there are no other physical length scales. As a result, the correlations do not depend on the distance between the points, since in the symmetric phase the distance does not exist. The distances appear only in the broken symmetry phase together with the geometry and with quantum mechanics.

In the same way the wave function in Eq.(\ref{CoulombNorm}) does not depend on coordinates when the symmetric state is approached (under condition that the electron mass $M_e$ multiplied by the fine structure constant $\alpha$ remains constant).
 
   \section{Open questions: from QFT to QM and then to CM}
   \label{open}
   
The correlations at infinity in the pre-geometric QFT state may be important for the interpretation of quantum mechanics, that emerges in the broken symmetry QM state. 

This includes:

Correlations between remote objects. The action at a distance, which violates principles of local causality, is the natural property of the pre-geometric symmetric phase, where the distance between points is absent. 

In the continuous limit, the quantum vacuum maintains the superluminal correlations of the pre-geometry, while matter emerging in the background of the vacuum is sub-luminal. The action at a distance for matter takes place via the superluminal vacuum.

Wave-particle duality emerges together with the Planck constants. Wave-particle duality is violated in both limit cases, in the QFT limit $1/\hbar \rightarrow 0$, and in the classical limit $\hbar/S \rightarrow 0$, where $S$ is the action for a single particle.

There are no closed systems. Due to correlations between far distant objects in the pre-geometric  quantum vacuum, all the material systems are open.

The problem of the collapse of the wave function can be considered through the prism of the pre-geometric QFT state. In the pure QFT (no distances, no relativistic particles, no mass, etc.) the wave function (or state vector) is not the physical object, since in QFT only the correlations are physical. The representation of correlations as bilinear combinations of wave functions is only a convenient tool to describe quantum mechanics. However, the wave function is an ill-defined quantity and is unobservable, since its phase is undefined. Only the phase difference and the frequency difference make sense. This is analogous to the introduction of a complex order parameter in a Bose condensate and in superconductors, which is a convenient tool for describing superfluidity and superconductivity. 
Here also only the phase difference and frequency difference have physical consequences. 
Also the wave functions for decaying particles and for quasiparticles are ill defined. 
 Since there is no physical wave function, its collapse does not make sense.

Decomposing correlations into a product of wave functions allows us to describe quantum mechanics using the Schrödinger equation. However, the Schrödinger equation is deterministic and therefore fails to describe the collapse of the wave function. The phenomenon of collapse reflects the randomness of the pre-geometric world, which is not deterministic at the pre-geometric level. The randomness is well described by quantum field theory, whose correlations only provide the probabilities of events, not the exact timing of them. Determinism develops in the process of coarse-graining, reaching complete determinism in the classical limit.  The irreversibility of the quantum collapse process is a consequence of the escape of some collapse product (e.g. photon) to infinity or beyond the event horizon.  Involving infinity in this process requires action at a distance.

 Weinberg proposes to consider the density matrix, rather than the state vector or wave function, as the physical reality,\cite{Weinberg2014} see also \cite{Hazeltine2025}. The evolution of the density matrix also reflects probabilities, not events. In this approach, the collapse can then be viewed as a quantum or thermodynamic fluctuation that transforms a mixed state described by density matrix into a less mixed state -- a state with lower entropy. It is the process by which off-shell (virtual) fluctuations of quantum fields hidden in the quantum vacuum leave the vacuum and become on-shell (real) if the vacuum state is disturbed or in the presence of excitations (i.e. matter). Example is the random quantum jumps in the process of decay of an unstable quantum system by spontaneous emission, see Ref. \cite{Raizen2025} and references therein. In the process of  $\alpha$-decay, a virtual alpha particle randomly jumps from one trajectory to another until it encounters an open trajectory leading into free space and becomes a real particle. 
 
Note that the particle trajectories emerge in the phase transition from QFT to QM, and these trajectories become classical in the CM limit. The integration over field variables in the QFT phase leads to a path integral formulation of QM, which in turn yields the laws of classical mechanics in the limit $\hbar\rightarrow 0$. The dominating role of QFT is in contrast to the canonical quantization approach in which quantum fields are obtained by quantizing classical fields, and which leads to the huge zero-point energy of the quantum vacuum. 

 Hawking radiation from the black hole by quantum tunneling and splitting of the black hole into smaller black and white holes by macroscopic quantum tunneling are the other examples of such quantum fluctuations. In both cases the entropy of the system decreases.\cite{Volovik2024} Similarly, entropy decreases after thermal fluctuations.
\cite{LL5} The black hole processes demonstrate that the black hole is a mixed state and it obeys thermodynamics with non-extensive Tsallis-Cirto entropy, $\sqrt{S(A+B)}= \sqrt{S(A)} +\sqrt{S(B)}$. 

The thermodynamic behaviour of quantum fluctuations suggests that the underlying pre-geometric quantum field theory may be based on some form of thermodynamics, the randomness of which leads to the appearance of mixed states in the emerging quantum mechanics. In other words, the quantum mechanical uncertainty may come from the hidden thermodynamics of the fundamental degrees of freedom of the quantum vacuum. The existence of a more fundamental theory that describes the randomness of the vacuum using a more fundamental and probably a more deterministic approach is an open question.

On the other hand, this randomness cannot be described using traditional zero-point fluctuations. The latter approach leads to the cosmological constant problem, among other serious conceptual difficulties including the inflationary scale which is far below the Planck energy scale. Even from the analogs of the quantum vacuum in condensed matter physics (the ground state of a many-body system), it is clear that quantum fluctuations cannot be viewed as a set of quantum oscillators in their ground states. Quantum fluctuations there are generated by microscopic (atomic) degrees of freedom. The thermodynamics of these "fundamental trans-Planckian" degrees of freedom naturally resolves the cosmological constant problem in Minkowski vacuum. The fundamental degrees of freedom destroy the present inflationary paradigm according to which the zero-point fluctuations of some field (inflaton) transform into density fluctuations at the end of inflation. This is the reason why this paradigm was challenged, see e.g. Ref. \cite{Perez2026} and references therein.

At the moment we know only the low-temperature properties of the quantum vacuum. The local entropy density of the de Sitter vacuum is\cite{Volovik2024}  
\begin{equation}
s(T)=\frac{3\pi}{4l^3_{\rm P}} \frac{T}{M_{\rm P}} \,,
\label{dSentropy}
\end{equation}
where $T=H/\pi$ is the local temperature of the de Sitter state. This temperature manifests itself only in the presence of matter: it is matter that perceives the de Sitter vacuum as a heat bath with this temperature. 
 Despite its thermal nature, the de Sitter state can still be considered a quantum vacuum. It contains no excitations of the vacuum -- no matter. The vacuum energy density -- the cosmological "constant" $\Lambda=sT/2$ -- represents the hidden quantum fluctuations. They obey the laws of thermodynamics, but only manifest themselves in the presence of matter, i.e. on-shell. The de Sitter thermal state of the quantum vacuum looks similar to the Zel'dovich stiff matter,\cite{Zeldovich1962} for which the speed of sound coincides with the speed of light.
 
In both the de Sitter vacuum and the black hole, the connection between quantum and thermal fluctuations is a consequence of event horizons -- the black hole horizon and the cosmological horizon, respectively. Here the horizon provides a pathway from off-shell to on-shell, and quantum fluctuations look as the thermal fluctuations in disguise. Whether in general the quantum fluctuations are the thermal fluctuations in disguise is an open question.  

Since the white holes and the contracting de Sitter demonstrate thermodynamics with negative entropy and negative temperature, it is not excluded that the fundamental microscopic degrees of freedom in the Diakonov approach may also obey the non-traditional thermodynamics with the generalized entropy and non-extensive statistics.

One of the predictions of the Diakonov theory is that there should be $N_F=4$ generations of fermions. This follows from  the 256 relevant degrees of freedom in the real spinor representations that unify quantum gravity and Standard Model.\cite{Diakonov2011}
This in particular suggests the extension of the  $G(224)$ left-right symmetric Pati-Salam model of the unification of quarks and leptons,  which is based on $SU(4)_C\times SU(2)_L\times SU(2)_R$ symmetry group,
 to the $G(22444)$ group:
 \begin{equation}
G(22444)=SU(4)_C\times SU(2)_L\times SU(2)_R \times SU(4)_F \times SO(3,1)\,.
\label{dSentropy}
\end{equation}
This extended group includes also the $SU(4)_F$ family group with $N_F=4$ generations and the $SO(3,1)$ Lorentz group, giving a total of $(2 + 2)\times 4\times 4\times 4=256$ real spinor degrees of freedom.
 The fourth generation neutrinos can be a good candidate for dark matter, in particular in the scenario of the so-called asymmetric dark matter (see e.g. Ref.\cite{Ishida2025} and references therein, and also Ref.\cite{Volovik2003a}). With four generations, one may suggest the microscopic length scale $a=l_{\rm UV}$ using Eq.(\ref{alpha}).

Diakonov approach also suggests that fundamental (pre-geometric) physics and hence the emerging quantum mechanics are based on real numbers. That the imaginary unit  may emerge together with relativistic fields and gravity was also obtained using the other scenarios of emergence:\cite{VolovikZubkov2014,VolovikZubkov2014a}  Ho\v{r}ava topological scenario of emergent Weyl fermions\cite{Horava2005} and Froggatt-Nielsen random dynamics theory\cite{FroggattNielsen1991}. This confirms that the imaginary unit is simply a product of the human mind, which becomes mathematically convenient in the emergent QM state.

  \section{Appendix}

 \subsection{Dimensions of Planck constants and their consequences}
 
  \subsubsection{Dimensions of Planck constants}

In conventional SI units the Planck constant is
 \begin{equation}
\hbar \approx 10^{-34} {\rm J}\cdot {\rm s}\,({\rm joule}\cdot{\rm second})\,. 
\label{hbardimension}
\end{equation}
In terms of the rest energy of gram, $M_{\rm g}={\rm g}  \cdot c^2$, the joule is
 \begin{equation}
{\rm J}={\rm kg}\cdot {\rm m}^2 \cdot {\rm s}^{-2}= \frac{10^3}{(3\cdot 10^8)^2} M_{\rm g}\approx10^{-14}M_{\rm g}\,,
\label{joule}
\end{equation}
and the Planck constant is
 \begin{equation}
\hbar \approx 10^{-48} M_{\rm g}\cdot {\rm s} \approx 10^{-43} M_{\rm P}\cdot {\rm s}\,. 
\label{hbarM}
\end{equation}

In the ADW approach, the mass after formation (mass of the formed Dirac particle or Planck mass $M_{\rm P}$) is a dimensionless quantity. That is why in this approach, $\hbar$ has dimension of time. If the Planck mass is chosen as the unit (the scale) of mass instead of the mass of cubic centimetre of water, then $\hbar$ becomes equal to the Planck time, and $\nh$ becomes the Planck length.
Such dimensions of Planck constants support the proposal  that $\hbar$ and $\nh$ enter the elements of the metric in Minkowski spacetime in Eq.(\ref{Minkowski}), which emerge in the ADW scenario and have dimensions $1/({\rm time})^2$ and $1/({\rm length})^2$.

How the Planck constants enter physical quantities, let us consider on example of quantum corrections to the metric.

 \subsubsection{Corrections to RN metric and dimensions}
 
The quantum corrections due to photon loops in the element $g_{00}$ of the Reissner-Nordstr\"om (RN) metric\cite{Donoghue2023,Petrov2016,Donoghue2002} can be written in following form: 
\begin{eqnarray}
\frac{g_{00}(r)} {g_{00}^{\rm Mink}}= \left(1 - 2\frac{M}{K} \frac{\nh}{r} + \frac{\alpha Q^2}{K} \frac{\nh^2}{r^2}\right) -\frac{8\pi}{3} \frac{\alpha Q^2}{MK} \frac{\nh^3}{r^3}\,.
\label{PhotonLoop}
\end{eqnarray}
where $K=M_{\rm Planck}^2$.

The 3 terms in brackets on the rhs of Eq.(\ref{PhotonLoop}) correspond to the conventional RN metric, where $Q$ is the electric charge of the black hole in units of the electric charge of electron, and 
$\alpha$ is the fine structure constant.
The quantum correction to the RN metric (the cubic term) changes the positions of two black hole horizons and thus the thermodynamics of the black hole.

Note that in terms of the Minkowski metric in Eq.(\ref{Minkowski}), the distance is $\tilde r = r/\nh$. It  has dimension of inverse mass $M$, i.e. $[\tilde r]=1/[M]$. Then one has
\begin{eqnarray}
\frac{g_{00}(\tilde r)} {g_{00}^{\rm Mink}}= \left(1 - 2\frac{M}{K\tilde r}+ \frac{\alpha Q^2}{K \tilde r^2} \right) -\frac{8\pi}{3} \frac{\alpha Q^2}{MK\tilde r^3} \,.
\label{PhotonLoop2}
\end{eqnarray}

To connect with Diakonov dimensions, the Eq.(\ref{PhotonLoop}) can be rewritten in the way, which demonstrates the dimensions of the parameters:
\begin{eqnarray}
\frac{g_{00}(r)} {g_{00}^{\rm Mink}}= \left(1 - 2\frac{M/M_0}{K/M_0^2} \, \frac{\nh/M_0}{r} + \frac{\alpha Q^2}{K/M_0^2}\, \frac{(\nh/M_0)^2}{r^2}\right) -
\nonumber
\\
-\frac{8\pi}{3} \frac{\alpha Q^2}{(M/M_0)(K/M_0^2)}\,  \frac{(\nh/M_0)^3}{r^3}\,.
\label{PhotonLoopRenormalized}
\end{eqnarray}
Here $M_0$ is the arbitrary mass, which represents the mass scale. This equation is invariant under the choice of $M_0$.
This equation renormalizes the Planck constant 
${\nh}$ to ${\nh}/M_0$, which has  the dimension of length. As a result the equation represents expansion in terms of ${\nh}/r$, which becomes dimensionless quantity if $K$ and $M$ become dimensionless.

  \subsubsection{Newton potential}

 The suggested Newton potential is:\cite{Shapiro2022,Frob2022,Khriplovich2004} 
 \begin{eqnarray}
U(r)= - \frac{M}{K} \frac{\nh}{r}\left[ 1+ C\, \frac{1}{K} \frac{\nh^2}{r^2} \right]\,,
\label{CorrectNewtonLaw}
\end{eqnarray}
where the dimensionless parameter $C$ contains contributions from 4 different fields in quantum vacuum:
 \begin{eqnarray}
C=\frac{41}{10\pi} + \frac{\frac{9}{4} N_0 + 6 N_{1/2} +12 N_1}{45\pi}\,.
\label{c}
\end{eqnarray}
Here the first term is the contribution from gravitons; $N_0$ is the number of scalars;
$N_{1/2}$ is the number of spinors; and $N_1$ is the number of gauge fields.
If the distance is expressed in terms of Minkowski metric in Eq.(\ref{Minkowski}), one has
 \begin{eqnarray}
U(\tilde r)= - \frac{M}{K\tilde r} \left[ 1+ C\, \frac{1}{K\tilde r^2}  \right]\,,
\label{CorrectNewtonLaw2}
\end{eqnarray}

With Eq.(\ref{CorrectNewtonLaw}) one can interpret gravity as quantum phenomenon, since in the classical limit ${\nh}\rightarrow 0$ the Newton interaction disappears.
The same is with Eq.(\ref{PhotonLoop}): the RN metric transforms to Minkowski.
This differs from the Donoghue arguments in favour of classical gravity,\cite{Donoghue2023} i.e. that the main term $-GM/r$ in the Newton potential $U(r)$ is classical.

 Eq.(\ref{CorrectNewtonLaw}) can be also rewritten in the renormalized from:
 \begin{eqnarray}
U(r)= - \frac{(M/M_0)}{(K/M_0^2)} \frac{(\nh/M_0)}{r}\left[ 1+ 
 \frac{C}{K/M_0^2} \frac{(\nh/M_0)^2}{r^2} \right]\,.
\label{NewtonLaw2}
\end{eqnarray}
As before, the parameter $M_0$ is the unity (scale) of mass, which can be arbitrarily chosen. When $M_0$ is chosen, all the masses and the gravitational coupling $K$ become dimensionless, while the normalized Planck constants ${\nh}/M_0$ and $\hbar/M_0$ acquire dimensions of length and time.

Eq.(\ref{CorrectNewtonLaw}) can be also written in terms of the variable $K(\tilde r)$:
 \begin{eqnarray}
K(\tilde r)= K_0\left(1 - \frac{C}{K_0 \tilde r^2}\right)\,.
\label{K}
\end{eqnarray}
This demonstrates that here we consider the limit $\tilde r \gg K_0^{-1/2}$, or $r \gg l_{\rm Planck}$ in conventional units.

The $1/r^2$ correction to $K$ in Eq. (\ref{K}) can be compared with the $T^2$ correction to the gravitational coupling,\cite{Volovik2003b}  
$K=K_0 -\tilde C T^2$. At the event horizon, $r=r_H$, the corresponding temperature is on the order of the Hawking temperature, $T\sim T_H=\nh/(4\pi r_H)$.
So, it is the same type of quantum correction, which leads to Hawking radiation. 

However, the Hawking radiation temperature in the main approximation is universal, i.e. it does not depend on the content of the quantum vacuum. There is the well known cancellation of different contributions: the Bekenstein-Hawking entropy is renormalized by the same quantum fluctuations in precisely the same way as the effective gravitational coupling $K$.\cite{Jacobson1994}

Such universality is lost in case of Newton law.

  \subsubsection{Bending of particles in gravitational field}
  
In the same way, the bending of massless particles in gravitational field of massive object\cite{Donoghue2023,HuanHangChi2019} can be written in terms of expansion $\nh/b$, where $b$ is the impact parameter:
\begin{eqnarray}
\theta= 4\frac{M}{K} \frac{\nh}{b} +\frac{15\pi}{4}\frac{M^2}{K^2} \frac{\nh^2}{b^2} 
+C_b\frac{M}{K^2} \frac{\nh^3}{b^3}\,,
\label{Bending}
\end{eqnarray}
or with impact parameter $\tilde b=b/\nh$ in the Minkowski metric in Eq.(\ref{Minkowski})
\begin{eqnarray}
\theta= 4\frac{M}{K\tilde b} +\frac{15\pi}{4}\frac{M^2}{K^2\tilde b^2} 
+C_b\frac{M}{K^2\tilde b^3} \,,
\label{Bending2}
\end{eqnarray}
or in the renormalized form
\begin{eqnarray}
\theta= 4\frac{M/M_0}{K/M_0^2} \, \frac{\nh/M_0}{b} +
\frac{15\pi}{4}\frac{M^2/M_0^2}{K^2/M_0^4} \, \frac{\nh^2/M_0^2}{b^2} +
\nonumber
\\
+C_b\frac{M/M_0}{K^2/M_0^4} \, \frac{\nh^3/M_0^3}{b^3}\,.
\label{BendingNormalization}
\end{eqnarray}

Here $C_b$ is the dimensionless parameter, which is different for scalars, photons and gravitons. $C_b$ also contains logarithm $\ln (r_0/b)$, where $r_0$ is an infrared cutoff.\cite{Donoghue2023,HuanHangChi2019}
Such cut-off can be provided by the Hubble parameter, i.e. by the cosmological horizon $r_0=r_c=\nh /(\hbar H)$, or by the Hawking temperature, $r_0=\nh/T$. 

The first two terms in the rhs of Eq.(\ref{Bending}) may suggest that there should be also another cubic term
 $\propto (M {\nh}/Kb)^3$.
We discuss this later.

 \subsubsection{Gravitational interaction between two masses}

The corrections to the gravitational interaction between two particles with masses $M_1$ and $M_2$ with distance $r$ between them are:\cite{Donoghue2023,Khriplovich2002,Khriplovich2004}  
\begin{eqnarray}
V(r)= - \frac{M_1M_2}{K} \frac{\nh}{r}\left[ 1+3\frac{M_1+M_2}{K} \frac{\nh}{r} 
+ \frac{41}{10\pi} \frac{1}{K} \frac{\nh^2}{r^2} \right],\,\,\,
\label{NewtonLaw}
\end{eqnarray}
or
\begin{eqnarray}
V(\tilde r)= - \frac{M_1M_2}{K\tilde r}  \left[ 1+3\frac{M_1+M_2}{K\tilde r}  
+ \frac{41}{10\pi} \frac{1}{K\tilde r^2}   \right].\,\,\,
\label{NewtonLaw4}
\end{eqnarray}

Here $\tilde r \gg M/K \gg K^{-1/2}$, or $r\gg r_H\gg l_{\rm Planck}$.

In the normalized form this is:
\begin{eqnarray}
V(r)= - \frac{M_1M_2/M_0^2}{K/M_0^2}\, \frac{\nh/M_0}{r} \times
\nonumber
\\
\left[ 1+3\frac{(M_1+M_2)/M_0}{K/M_0^2}\, \frac{\nh/M_0}{r} 
+ \frac{41}{10\pi} \frac{1}{K/M_0^2} \,\frac{(\nh/M_0)^2}{r^2} \right].\,\,\,\,
\label{NewtonLawNormalized}
\end{eqnarray}

Eq.(\ref{NewtonLaw}) suggests that the Newton attraction, $-GM_1M_2/r$, which is characterized by the Newton constant $G=\nh/K$, is the quantum phenomenon. 

The same takes place with Coulomb interaction $e^2/r=  (\nh/r) \alpha$, which is also the quantum phenomenon. This interaction appears due to the breaking of the double Lorentz symmetry in the quantum vacuum, $L\times L \rightarrow L$. In this approach, the gravitational tetrads $e^a_\mu$ emerge as the order parameter of the phase transition.\cite{Volovik1990,Diakonov2011,Diakonov2012,Volovik2022} 
At this transition, the metric also appears, which in turn gives rise  to the $FF/\alpha$ term in the action for the electromagnetic field and thus to the Coulomb interaction. That is why the emerging Coulomb interaction represents the quantum field effect.

The same concerns the $K{\cal R}$ term in the gravitational action. But now the Newton constant $G=\nh/K$ appears due to the further symmetry breaking in the quantum vacuum: the breaking of the conformal (scale) symmetry.

In Eq.(\ref{CorrectNewtonLaw}) the extra cubic term $\propto (M\nh/Kr)^3$ is also missing. Maybe it does not exist due to special reasons,\cite{Kirilin2007} i.e. "the quantum corrections to metric must be independent of the choice of
field variables, i.e. must be reparametrization invariant". For example,
the combination of the nonlinear terms can be transformed to the linear term:
\begin{eqnarray}
 a_1\frac{M}{K} \frac{\nh}{r} + a_2 \frac{M^2}{K^2} \frac{\nh^2}{r^2} + 
 a_3 \frac{M^3}{K^3} \frac{\nh^3}{r^3} + ... = a_1 \frac{M}{K} \frac{\nh}{r'}  \,,
\label{reparametrization}
\end{eqnarray}
or
\begin{eqnarray}
r'= r \left( 1 - \frac{a_2}{a_1} \frac{M}{K} \frac{\nh}{r} + ...\right) \,,
\label{reparametrization2}
\end{eqnarray}

 The same reason does not allow also the quadratic term in Eq.(\ref{NewtonLaw}), because  the similar quadratic term $-6(M/K)^2(\nh/r)^2$ does not contribute to the Eq.(\ref{PhotonLoop}) for the black hole metric. Near the black hole horizon, such nonlinear (quadratic, cubic, etc.) terms are on the order of the linear term, which certainly contradicts to the GR.
 
 On the other hand the cubic quantum corrections in Eqs. (\ref{Bending}) and (\ref{NewtonLaw}) are physical. They have the same origin as the $T^2$ temperature correction to the gravitational coupling $K$, which is valid even for effective gravity in condensed matter.\cite{Volovik2003}

\subsubsection{Classical dynamics of a point particle}

Another example is the dimensionless action for the classical dynamics of a point particle in Minkowski spacetime in ADM approach:
  \begin{eqnarray}
S_{\rm ADW}=M\int ds=M\int \frac{dt}{\hbar} \sqrt{1 - \frac{v^2}{c^2}} =
\label{Classicalparticle1}
\\
= \frac{(M/M_0)}{(\hbar/M_0)}\int dt \sqrt{1 - \frac{v^2}{c^2}}\,.
\label{Classicalparticle2}
\end{eqnarray}
Again, when mass becomes dimensionless, the normalized Planck constant $\hbar/M_0$ acquires the dimension of time. 

Any action in ADM scenario must be dimensionless, since in the symmetric phase there are no Planck constants $\hbar$ and $\nh$. They appear only in the broken symmetry state. That is why the action can play the role of the phase in $e^{i S_{\rm ADW}}$.

 \subsubsection{Wave function of electron in Coulomb field}
 
The competition of limits can be also seen in the bra-ket Dirac notation. The overlap of the quantum states localized at points ${\bf r}_2$ and ${\bf r}_1$ is
 \begin{equation}
<{\bf r}_1|{\bf r}_2> = \frac{1}{\sqrt{\gamma}} \delta({\bf r}_2-{\bf r}_1)=\nh^3  \delta({\bf r}_2-{\bf r}_1) \,.
\label{localized1}
\end{equation}
Here $\gamma=1/\nh^6$ is the determinant of the space part of the Minkowski metric.
We have
\begin{equation}
|\Psi>=\int d^3 r \sqrt{\gamma}\Psi({\bf r)} |{\bf r}>\,.
\label{Psi}
\end{equation}
and
\begin{eqnarray}
<\Psi|\Psi>=\int d^3 r d^3 r' \gamma \Psi^*({\bf r}')\Psi({\bf r)}<{\bf r}|{\bf r}'>=
\nonumber
\\
= \int d^3 r  \sqrt{\gamma}|\Psi|^2= \int \frac{d^3 r }{\nh^3} |\Psi|^2=1\,.
\label{PsiSquare}
\end{eqnarray}

Let us consider the wave function of electron in Coulomb field. With traditional metric, i.e. for $\gamma=1$ it is:
\begin{equation}
|\Psi|^2= \frac{1}{\pi a^3}\exp(- 2r/a) \,\,,\,\, a=\frac{\nh}{M_e\alpha}\,.
\label{Coulomb}
\end{equation}
With $\gamma=1/\nh^6$, one obtains from Eq.(\ref{PsiSquare}):
\begin{equation}
|\Psi|^2= \frac{(M_e\alpha)^3}{\pi}\exp(- 2M_e\alpha r/\nh)  \,.
\label{CoulombNorm}
\end{equation}
 
It again depends on which limit is taken first:
\begin{equation}
\Psi(r \rightarrow \infty)=0 \,\,,\,\, \Psi(\nh \rightarrow \infty)=(M_e\alpha)^{3/2}/\sqrt{\pi}\,. 
\label{ComperitionOfLimits}
\end{equation}
In the limit $1/\nh \rightarrow 0$ the wave function is constant in space (if $\alpha M_e$ is kept constant).
 
 On the other hand, instead of Eq.(\ref{ComperitionOfLimits}) one can write:
 \begin{equation}
\Psi(r)_{\nh \rightarrow 0}=0 \,\,,\,\, \Psi(r)_{\nh \rightarrow \infty}={\rm constant}\,. 
\label{ComperitionOfLimits2}
\end{equation}
This demonstrates that quantum mechanics disappears in both limits: In the limit $\nh \rightarrow 0$ one has classical mechanics, while in the limit $1/\nh \rightarrow 0$ one has quantum field theory.

\subsection{Quantum jumps on micro-level and choice of the final state on macro-level}
\subsubsection{Vortex nucleation by neutron irradiation}

\begin{figure}[htt]
 \includegraphics[width=\columnwidth]{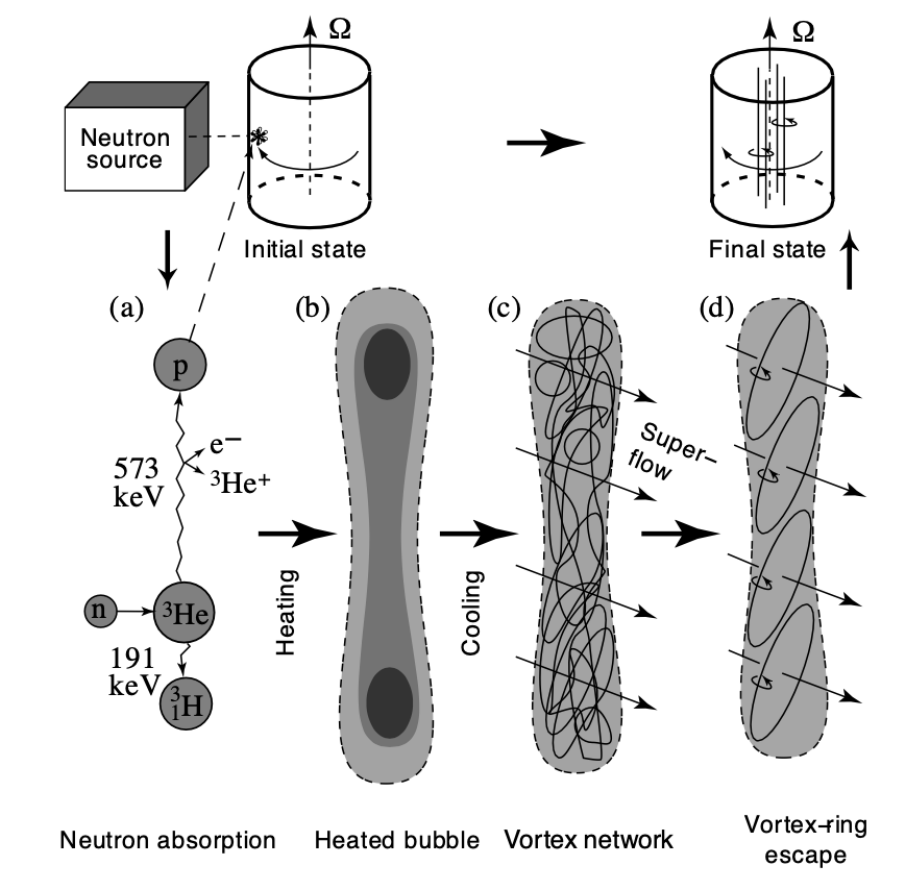}
 \caption{Illustration of vortex formation after reaction $n$ + $^3$He $\rightarrow$ $^3$H + $^1$H + 0.764 MeV, from Ref.\cite{Eltsov2005}}
 \label{fissionFig}
\end{figure}

The quantum jumps in the spontaneous emission,\cite{Cook1985,Raizen2025} is one of the interesting topics in quantum field theory and quantum mechanics.
Here we discuss experiments on creation of the topological defects in superfluid $^3$He by the Kibble-Zurek mechanism. \cite{Kibble1996,Kibble2000,Eltsov2005}  
In this experiment, the superfluid $^3$He in cylindrical container is irradiated by neutrons, see Fig. \ref{fissionFig}.
 The emitted neutron produces the fission reaction, and the released kinetic energy warms a small regions of the liquid. The further cooling down of these regions through the broken symmetry phase transition produces quantized vortices in the Kibble-Zurek mechanism. After that the created vortices become concentrated in the central region of the container forming the vortex cluster. 
 
 \begin{figure}[htt]
 \includegraphics[width=\columnwidth]{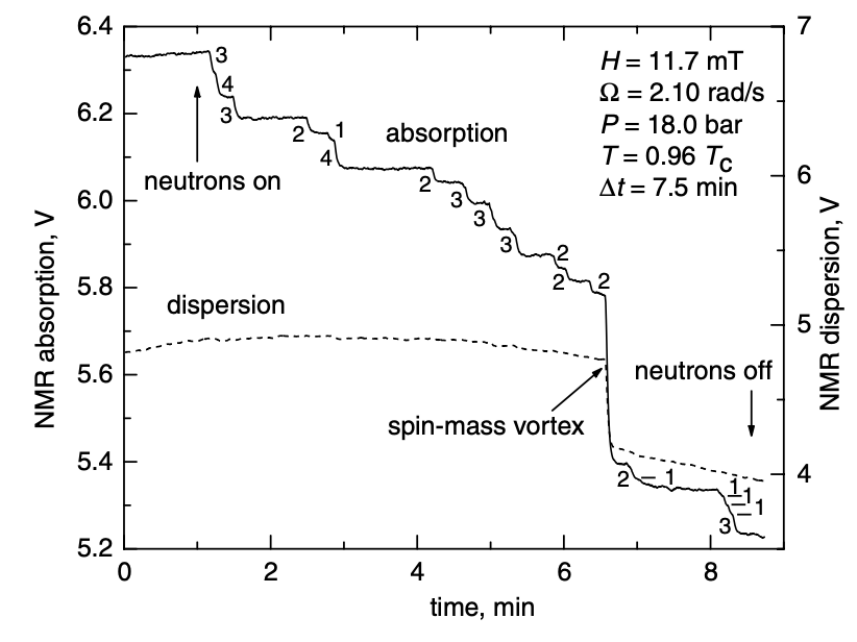}
 \caption{
 Neutron-irradiation record of $^3$He-B. Each of the small downward steps in the absorption record marks a neutron absorption event and its height measures the number of newly formed vortex lines. The single large step is attributed to the creation of spin-mass vortex with the soliton tail.
 From Ref. \cite{Kibble2000} }
 \label{Figure1}
\end{figure}

 \begin{figure}[htt]
 \includegraphics[width=\columnwidth]{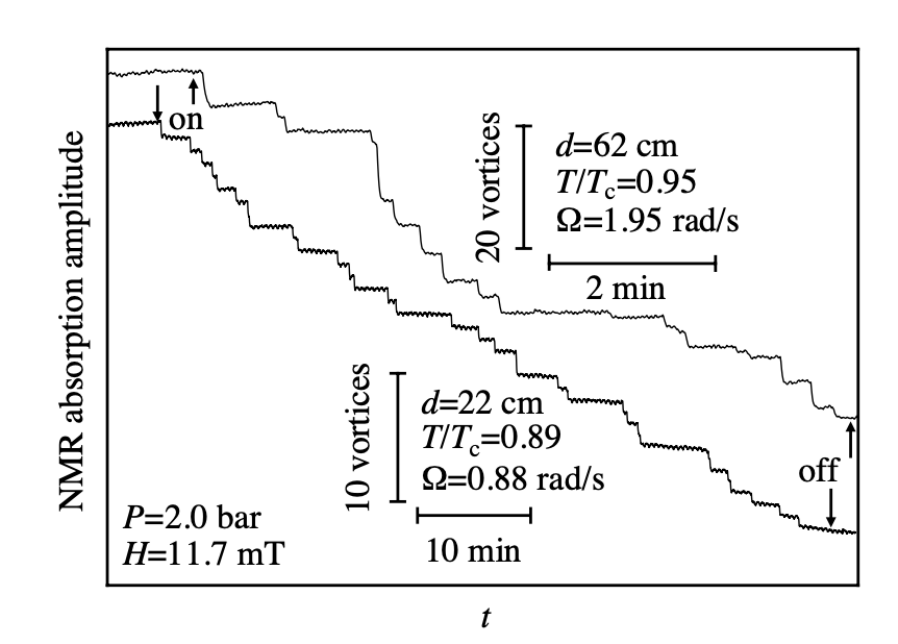}
 \caption{ Neutron-irradiation record  for two angular velocities $\Omega$ of rotating cryostat. From Ref. \cite{Eltsov2005} }
 \label{Figure2}
\end{figure}

The experiment is continued  without the resetting.  The  nuclear reaction, which causes the Kibble-Zurek effect, takes place in the region close to the walls of container (the neutron mean free path is about 100 $\mu$m. It is not influenced by the already created vortices concentrated in the central region. After each new fission event the number of vortices in the vortex cluster increases step by step, which is measured using the NMR techniques, see Figs. \ref{Figure1} and  \ref{Figure2}. Each random jump is the result of the creation of macroscopic quantized vortices by a neutron entering the liquid. The number of vortices created in one event is random.  The original metastable state (the vortex-less state in the rotating container) goes by jump to another metastable state with few vortices  concentrated in the central region of the rotating container. When the next neutron enters the liquid, there is the jump to the further metastable state with larger number of vortex lines in the vortex cluster, and so on. The process continues until the stable state with the equilibrium number of vortices is reached, where the full container is occupied by vortices.

\subsubsection{Quantum and classical randomness}

The jumps occur randomly and the number of the vortices created at a single jump is also random. The average number of vortices created at a single jump depends on temperature and the angular velocity $\Omega$ of rotating cryostat, see Fig. \ref{Figure2}. The large jump in Fig. \ref{Figure1} is the result of the formation of the composite object: topological solition terminated by spin-mass vortex. However, the randomness in the time of jump and the randomness in the number of created vortices have different origin,  quantum and classical correspondingly, although quantum mechanics (or relativistic quantum field theory) is behind all the processes.

Quantum mechanics or relativistic quantum field theory (RQFT) in general properly describes each step in this process:
 
1) Neutrons are produced by quantum mechanical spontaneous fission.

2) Neutron itself is the composite object --  the product of quantum chromodynamics (QCD).

3) The nuclear fission in Fig. \ref{fissionFig}  is described by the relativistic quantum field theory.
 
4) Superfluid $^3$He is the quantum liquid. 

5) Vortex in superfluid $^3$He is of the quantum origin: its circulation is quantized in units  $\pi\hbar/m_4$, where $m_4$ is the mass of helium atom. The spin vortex and soliton are also quantized topological objects. 

While quantum mechanics (or RQFT) correctly describes everything, the whole process of measurement can be divided in two parts: 
(1)  the neutron radiation and nuclear reaction occurring on microscopic level; and 
(2) processes which are described on the macroscopic level (warming, cooling, Kibble-Zurek process of the vortex creation, relaxation to the quasi-equilibrium state).  
 The observed randomness of the results of the experiments come from both levels.
 The randomness in the number of vortices created in each event and the randomness in the type of the created topological objects come from the macroscopic statistical mechanics. 
On the other hand, the discrete jumps  in the radioactive decay of unstable nuclei  and the randomness of these jumps are the properties of the microscopic physics. Such discrete quantum jump is usually associated with the so-called wave function collapse, although it still remains unclear whether the wave function is a well-defined quantity.

The initial vortex network forms during cooling through $T_c$ with the relevant time scale of microseconds. The later evolution of the network to form vortex cluster takes time from milliseconds up to seconds. These two time scales are essentially shorter than the interval between the jumps, which is determined by the intensity of neutron radiation and is about a minute. That is why the jump in the number of vortices is the experimental demonstration of the quantum jump.

\subsubsection{Origin of quantum jumps}

While quantum mechanics works in both levels, micro and macro, we still do not know the reason of the randomness at the micro-level, i.e. why the micro-level is described only by the probabilities of the events. Stochastic behavior of microscopic systems is an observational fact established beyond any reasonable doubt,\cite{Frohlich2024} 
and the unpredictability of the quantum jumps at this level is the main unsolved problem of quantum field theory. It is clear from the above experiments that the fundamental randomness at the micro-level has no relation to the measurement process, which is practically governed by macro-physics, and correspondingly  has no relation to observer who sees the twenty years old data of detectors in Figs. \ref{Figure1} and  \ref{Figure2}.

Comparison with the statistical physics at macro-level may suggest the scenario in which the physics at the micro-level can be also regulated by the hidden macroscopics. The virtual (off-shell) quantum vacuum is turbulent (stochastic), while the vacuum is static on-shell, and its stochastic nature is hidden. The randomness of the quantum vacuum manifests itself when the vacuum is excited, and its turbulent quantum fluctuations leak out from the vacuum and become real.  So, in principle, any violation of the equilibrium vacuum state opens the channel for the virtual to become real. Event horizon may serve as one of the channels. It is possible that all types of excitations of the vacuum (particles and fields) eventually decay, although the life time of some states can be rather long. In other words, there are no stable particles. 
Although this scenario does not explain the physics of a quantum jump, the randomness of the quantum vacuum is the perspective direction.

Quantum mechanics, viewed in terms of path integrals, suggests that particles try one trajectory after another until they find one that opens a path out. This is the moment of a quantum jump in the stochastic decay of unstable nuclei. This jump leads to the creation of neutron, which in turn triggers the quantum jump in the process of nuclear fission, and finally this leads to the macroscopic process of vortex formation.

    \subsection{Gravitational field as specific matter field}
    \label{GravFieldSection}

Since in Diakonov  pre-geometric theory the gravitational field originates from the quantum matter fields, it is natural to treat the gravitational field as any other matter field such as electromagnetic field, i.e. gravity is also the part of matter. 
Then the conventional Einstein–Hilbert action can be rewritten in the following form:
 \begin{equation}
S= \int d^4x \, {\cal L}^{\rm m}+ \int d^4x \, {\cal L}^{\rm grav}   \,\,,\,\, 
 {\cal L}^{\rm grav} =\frac{\sqrt{-g}}{16\pi G} \,g^{\mu\nu} {\cal R}_{\mu\nu} \,.
\label{StandardAction}
\end{equation}
where ${\cal R}_{\mu\nu}$ is the Ricci curvature tensor. The Lagrangian of the ordinary matter fields is ${\cal L}^{\rm m}=\sqrt{-g}L^{\rm m}$, where $L^{\rm m}$ is the traditional Lagrangian. This form of the Lagrangian is natural for the ADW theory,  where the original action (\ref{OriginalActions}) has no normalization of the 4-volume by the metric determinant, because the metric emerges only at low energy.
  
With this form of the ${\cal L}^{\rm grav}$, the Einstein equations, which are usually obtained  from the functional derivative of the action, $\delta S/\delta g^{\mu\nu}=0$, can be obtained
 from the conventional derivative of the Lagrangian, $\frac{d{\cal L}}{d g^{\mu\nu}} =0$.
 The reason is that the variation of the Ricci tensor ${\cal R}_{\mu\nu}$ leads to the total derivative, and can be ignored with the proper choice of the boundary conditions. The same takes place for the matter action.

That is why the energy-momentum tensor of the gravitational field, as well as the energy-momentum tensor  of any other matter filed, can be obtained as the derivative of the corresponding Lagrangian  with respect to the metric:
 \begin{equation}
T^{\rm m}_{\mu\nu}=-\frac{2}{\sqrt{-g}} \frac{d{\cal L}^{\rm m}}{d g^{\mu\nu}}
  \,\,,\,\, 
T^{\rm grav}_{\mu\nu}=-\frac{2}{\sqrt{-g}} \frac{d{\cal L}^{\rm grav}}{d g^{\mu\nu}}
=-\frac{1}{8\pi G} \left(  {\cal R}_{\mu\nu} -\frac{1}{2} {\cal R}g_{\mu\nu}\right)
 \,.
\label{EnergyMomentum}
\end{equation}
Then the Einstein equation $\frac{d{\cal L}}{d g^{\mu\nu}} =0$ simply means that the total energy-momentum tensor of the system (which includes ordinary matter,  the "matter of gravity" and also dark matter and dark energy) is  equal to zero:
\begin{equation}
T_{\mu\nu}=T^{\rm m}_{\mu\nu} + T^{\rm grav}_{\mu\nu}+ T^{\rm dark}_{\mu\nu}=0
\,.
\label{ZeroGeneralized}
\end{equation}
Being zero, the total energy-momentum tensor of the Universe $T_{\mu\nu}$ is automatically conserved.
Eq.(\ref{ZeroGeneralized}) may support the idea of zero-energy Universe,\cite{Rosen1994} the zero Hamiltonian in Arnowitt-Deser-Misner (ADM) formalism,\cite{Nunez2022} and the Universe from Nothing.\cite{KraussBook}

\end{document}